\documentstyle{article}
\begin{document}
\begin{sloppypar}
\newcommand{\be}{\begin{equation}}
\newcommand{\ee}{\end{equation}}
\large
\begin{center}
{\bf Dielectric properties of multiband electron systems: 

I - Tight-binding formulation}

\vspace{5mm}

 P. \v{Z}upanovi\'{c}
 
{\em Department of Physics,  
Faculty of Science and Art, University of Split,  
Teslina 10, 21000 Split, Croatia} \vspace{5mm}\\
 A. Bjeli\v{s} and  S. Bari\v{s}i\'{c}

{\em Department of Physics,  
Faculty of Science, University of Zagreb,  
P.O.B. 162, 

10001 Zagreb, Croatia}
\end{center}
\vspace{5mm}

\begin{center}
{\bf Abstract}
\end{center}
\mbox{} 

The screened electron-electron interaction in a multi-band electron system is 
calculated within the random phase approximation and in the tight-binding
representation. The obtained dielectric matrix contains, beside the usual
site-site correlations, also the site-bond and bond-bond correlations, and
thus includes all physically relevant polarization processes. The arguments
are given that the bond contributions are negligible in the long wavelength
limit. We analyse the system with two non-overlapping bands in this limit,
and show that the corresponding dielectric matrix reduces to a $2\times2$
form. The intra-band and inter-band contributions are represented by diagonal
matrix elements, while the off-diagonal elements contain the mixing between
them. The latter is absent in insulators but may be finite in conductors.
Performing the multipole expansion of the bare long-range interaction, we
show that this mixing is directly related to the symmetry of the atomic
orbitals participating in the tight-binding electronic states. In systems
with forbidden atomic dipolar transitions, the intra-band and inter-band 
polarizations are separated. However, when the dipolar transitions are
allowed, the off-diagonal elements of the dielectric matrix are of the same
order as diagonal ones, due to a finite monopole-dipole interaction between
the intra-band and inter-band charge fluctuations. We also calculate the
macroscopic dielectric function and obtain an expression which interpolates
between the well-known limits of one-band conductors and pure insulators.
In particular, it is shown that the microscopic origin of the so-called
self-polarization corrections is the on-site interaction which exchanges
two electrons at different orbitals, combined with a finite tunneling
between neighboring sites. 

\bigskip


\bigskip

   
\newpage  
\section {Introduction}

The present and the forthcoming \cite{zbbII} work revisit the problem of
the dielectric properties of multiband metals and insulators. Early
calculations \cite{nozieres} were carried out neglecting the discreteness
of the lattice in the derivation of the screened Coulomb interaction. The
result was a dielectric function built adding the intraband and interband
polarizabilities [2 - 4]. The continuous approximation in calculating the
Coulomb matrix elements is however inconsistent with the multiband
assumption, since the latter implies the existence of the lattice. Later
works attempted to include properly the local electric field effects
associated to the lattice discreteness, by formulating the problem of
dielectric response in the representation of the {\bf k+G} plane waves,
where {\bf G} are vectors of the reciprocal lattice \cite{adler,wiser}.
In this representation one comes to the dielectric matrix
$\varepsilon(\mbox{\bf q+G, q+G'}, \omega)$ which is of infinite order for
a given {\bf q}. Since this latter is very difficult to handle in physical
terms, it was necessary either to make additional approximations [7 - 12]
(beside the initial random phase approximation (RPA)), or to develop
various numerical  procedures [13 - 15], which in turn usually do not
provide a simple physical interpretation of the results obtained.  

A particular method in this direction is based on the use of the
tight-binding (TB) basis in the calculation of the polarizabilities which
figure in the matrix elements of $\varepsilon(\mbox{\bf q+G, q+G'}, \omega)$
[8 - 12]. Providing that the atomic orbitals which represent this basis are
localized enough, one gets to a good approximation a factorizable form of
dielectric matrix elements, introduced earlier on through the so-called
generalized shell model by Sinha et al \cite{sinha}. This factorization
in principle enabled the analytical inversion of the dielectric matrix,
and led to significant improvements in the understanding of dielectric
properties of simple TB band structures \cite{hash,giaq}. The whole
approach was still burdened by the unnecessary parallel use of two basis,
i.e. plane waves and TB states.

The aim of the present paper is to show that the problem of the dielectric
response is significantly simplified when the TB scheme is used from the
outset, in the definition of the dielectric matrix itself. The TB approach
for the single band case was developed before \cite{barisic}. In the
multiband case the elements of the TB dielectric matrix have a transparent
physical content, i.e. they are defined by the pairs of intraband and/or
interband transitions and by an index which covers pure site-site
correlations as well as the site-bond and bond-bond correlations between
neighboring lattice points. By taking into account these bond
correlations we complete our earlier treatment of the dielectric response
in the TB approach \cite{zbb}. Since the atomic orbitals which build the
TB Bloch functions are well localized (in contrast to usual Wannier
functions), it is well justified to keep only the bond contributions between
first neighbors, so that the index mentioned above is of the order of a
coordination number for a given lattice.
After an additional truncation by keeping a finite number of presumably 
relevant electronic bands, the dielectric matrix reduces to a finite,
rather low order. 

In order to illustrate the method, we consider here in some detail 
the simple case of two  non-overlapping electron bands, allowing for two
types of symmetry of the TB orbitals which form the empty higher
(conducting) band. In all cases the lower band is assumed to be partially
or fully filled. We calculate all screened TB Coulomb matrix
elements within RPA, and show that in the long-wavelength limit the 
problem reduces to a $2\times2$ dielectric matrix. The same result was
obtained by an alternative analysis via the Heisenberg equations for the
electronic charge density \cite{zbb}. Since the present approach gives in
addition screened Coulomb matrix elements for all intraband and interband
processes, it can be conveniently used in the calculation of the screened
local fields, such as those produced by phonons in ionic multiband metals,
and in particular in high $\mbox{T}_{c}$ superconductors.

Since the range of the TB atomic (molecular) orbitals is shorter than that
of the two-site Coulomb interactions \cite{kohn}, the latter can be 
expanded into the multipole series, which is particulrly well controlled
in the long wavelength limit. In this limit one recognizes the decisive
role of the symmetry of TB orbitals in the multiband dielectric screening.
When two orbitals have  the same parity, the intra-band and inter-band
polarization processes are decoupled, so that the total dielectric function
is the product of the corresponding intra-band and inter-band dielectric
functions [19 - 21]. A more interesting situation occurs when the band
symmetry allows for finite dipolar transitions. The finiteness of the
monopole-dipole coupling in this case leads to qualitatively different
results for the microscopic dielectric function (defined as the determinant
of the dielectric matrix) and the TB matrix elements of screened
electron-electron interaction. The effects of this coupling on the
collective modes of the two-band electron liquid will be analysed in the
paper II \cite{zbbII}. 

The multipole expansion also separates the long-range part of the bare
Coulomb interaction from the on-site contributions. Because of that, we
are able to trace the microscopic origin of the so-called self-polarization
corrections \cite{adler} to the macroscopic dielectric function
$\epsilon_{M}({\bf q},\omega)$, which can be easily determined once the
TB matrix elements of the screened Coulomb interaction are known.

The paper is organized in the following way. In Sect.2  we derive the linear
system of RPA equations for the screened Coulomb interaction and introduce
the corresponding dielectric matrix.  In Sects.3 to 5 we apply these general
results to the two-band model. The explicit expressions for the microscopic
dielectric function and all matrix elements of the screened interaction are
derived in Sect.3. In Sect.4 we perform the multipole expansion and
discuss the role of the short-range and long-range contributions to the
microscopic dielectric function. The macroscopic dielectric function and the
origin of the self-polarization corrections are considered in Sect.5. Sect.6
contains some concluding remarks. 

\section {The tight-binding formulation of the dielectric matrix}

The Dyson's equation for the screened  interaction  in the direct space
is within RPA given by
\begin{equation}
\label{a1}
\overline{V}({\bf r},{\bf r'},\omega)=V({\bf r}-{\bf r'})+
                     \int d{\bf r}_{1}
\int d{\bf r_{1}^{'}} \,V({\bf r}-{\bf r_{1}})
 \Pi({\bf r_{1}},{\bf r_{1}^{'}},\omega)
\overline{V}({\bf r_{1}^{'}},{\bf r^{'}},\omega),
\end{equation}
where 
\begin{equation}
\label{a2}
\Pi({\bf r},{\bf r^{'}},\omega)=-\frac{i}{\pi}
 \int d \omega^{'} \, G^{0}({\bf r},{\bf r^{'}},\omega+\omega^{'})
G^{0}({\bf r^{'}},{\bf r},\omega^{'})
\end{equation}
is the bubble polarization diagram, and $G^{0}({\bf r},{\bf r^{'}},\omega)=
\sum_{l}G_{l}^{0}({\bf r},{\bf r^{'}},\omega) $ is a bare
Green's function with the contribution from the $l$-th band given by
\begin{eqnarray}
\label{a3}
G_{l}^{0}({\bf r},{\bf r^{'}},\omega)
 &=&\sum_{\bf k} \psi_{l,{\bf k}}({\bf r}) \psi^{*}_{l,{\bf k}}({\bf r^{'}})
 \times \nonumber \\
 && \times\left\{ \frac{1-n_{l}({\bf k})}{\omega-E_{l}({\bf k})+i\eta} 
+\frac{n_{l}({\bf k}) }
{ \omega-E_{l}({\bf k})-i\eta } \right\},
\end{eqnarray}
with $\psi_{l,{\bf k}}({\bf r})$, $E_{l}({\bf k})$ and $n_{l}$ being the
corresponding Bloch function, band dispersion and occupation number
respectively.  Our aim is to transform the integral equation (\ref{a1})
into a system of algebraic equations for the TB matrix elements of the
screened potential. To this end we recall the essential property of the TB
basis, namely the weak overlap of the atomic orbitals at neighboring
crystal sites. Thus, it suffices to keep in various matrix elements only
the contributions with pairs of orbitals centered at the same and
neighboring sites, and to neglect all contributions with more distant
pairs of orbitals. In this respect it is important to compare the TB and
the Wannier bases. Although the Wannier functions at different sites are
exactly orthogonal, they usually have slowly converging oscillatory 
tails, which greatly complicates the evaluation of the one- and two-body
matrix elements. We therefore continue by strictly using the TB basis.
The Bloch functions are thus given by
\begin{equation}
\label{a4}
\psi_{l,{\bf k}}({\bf r}) = \frac{1}{\sqrt{N}} 
\frac{1}{[1 + S_l({\bf k})]^{1/2}}  \sum_{\bf R} e^{i{\bf k R}} 
\varphi_l({\bf r - R}).
\end{equation}
Here
\begin{equation}
\label{a5}
S_l({\bf k}) = \sum_{\mbox{\boldmath $\delta$} \neq {\bf 0}} e^{i{\bf k}
\mbox{\boldmath  $\delta$}}
 S_{l}(\mbox{\boldmath $\delta$}),
\end{equation}
where 
\begin{equation}
\label{a6}
S_{l}(\mbox{\boldmath $\delta$}) =
 \int d{\bf r} \varphi^{*}_l({\bf r}) \varphi_l({\bf r}
 - \mbox{\boldmath $\delta$}) 
\end{equation}
is the direct overlap between nearest neighboring atomic orbitals
$\varphi_l$, and the sum in (\ref{a5}) involves only first neighbors.
Although the further considerations mainly do not depend on the details
of the band dispersions $E_{l}({\bf k})$, we remind that they include,
beside $S_l({\bf k})$, also sums of tunneling integrals due to the effective
ionic potential. In these sums it again suffices to keep only nearest
neighbors. 

Let us now multiply eq.(\ref{a1}) by $e^{i{\bf q}({\bf R} + 
\frac{{\bf \delta}_1 - {\bf \delta}_2}{2})}\varphi_{l_{1}}^{*}
({\bf r}-{\bf R} -\mbox{\boldmath $\delta$}_1)\varphi_{l_{2}}^{*}({\bf r'})
\varphi_{l_{1}^{'}}({\bf r}-{\bf R})\varphi_{l_{2}^{'}}({\bf r'}
-\mbox{\boldmath $\delta$}_2)$, 
where ${\bf R}$, {\boldmath $\delta$}$_1$  and {\boldmath $\delta$}$_2$
denote crystal sites. After the integration with respect to ${\bf r}$ and
${\bf r'}$ and summation with respect to ${\bf R}$ one gets the 
linear system of equations
\begin{eqnarray}
\label{a7}
\sum_{l_{3}l_{3}^{'}} \sum_{\mbox{\boldmath $\delta$}_{3}} 
\left[ \delta_{l'_{3}l_{1}}
 \delta_{l_{3}l'_{1}}  \delta_{\mbox{\boldmath $\delta$}_{3}
\mbox{\boldmath $\delta$}_{1}} - \sum_{\mbox{\boldmath $\delta$}'_{3}} 
V_{l_{1}l_{3}l_{1}^{'}l'_{3}}
(\mbox{\boldmath $\delta$}_1, \mbox{\boldmath $\delta$}'_3; {\bf q})
\Pi_{l_{3}l_{3}^{'}}(\mbox{\boldmath $\delta$}'_{3}-
\mbox{\boldmath $\delta$}_{3},{\bf q},\omega) \right] 
\times \nonumber\\
\times
\bar{V}_{l'_{3}l_{2}l_{3}l'_{2}}(\mbox{\boldmath $\delta$}_3,
 \mbox{\boldmath $\delta$}_2; {\bf q},\omega)=
V_{l_{1}l_{2}l_{1}^{'}l'_{2}}(\mbox{\boldmath $\delta$}_1, 
\mbox{\boldmath $\delta$}_2; {\bf q})
\end{eqnarray}
for the matrix elements of the screened Coulomb interaction
\begin{eqnarray}
\label{a8}
\bar{V}_{l_{1}l_{2}l_{1}^{'}l'_{2}}(\mbox{\boldmath  $\delta$}_1,
\mbox{\boldmath $\delta$}_2; {\bf q},\omega)= 
 \sum_{{\bf R}}e^{i{\bf q}({\bf R} + 
\frac{\mbox{\boldmath  $\delta$}_1 - \mbox{\boldmath  $\delta$}_2}{2})}  
 \int d{\bf r} \int d{\bf r'}
\varphi_{l_{1}}^{*}({\bf r}-{\bf R}-\mbox{\boldmath $\delta$}_1)	 
 \varphi_{l_{2}}^{*}({\bf r'}) \times \nonumber \\
 \times \bar{V}({\bf r},{\bf r'},\omega)
\varphi_{l_{1}^{'}}({\bf r}-{\bf R})\varphi_{l_{2}^{'}}({\bf r'}-
\mbox{\boldmath $\delta$}_2).
\end{eqnarray}
$V_{l_{1}l_{2}l_{1}^{'}l_{2}^{'}}(\mbox{\boldmath $\delta$}_1, 
\mbox{\boldmath $\delta$}_2; {\bf q})$ are the corresponding 
bare matrix elements, when $\bar{V}({\bf r},{\bf r'},\omega)$ in (\ref{a8})
is replaced by $e^2/|{\bf r} - {\bf r'}|$. In this TB representation of
two-body interaction one again relies on the localized nature of atomic
orbitals and retains only those matrix elements in which
$\mbox{\boldmath $\delta$}_1$ and $\mbox{\boldmath $\delta$}_2$ are either
zero or denote nearest neighboring sites. In other words, one distinguishes
between the site-site ($\mbox{\boldmath $\delta$}_1 = 
\mbox{\boldmath $\delta$}_2 = 0$), site-bond
($\mbox{\boldmath $\delta$}_1$ or $\mbox{\boldmath $\delta$}_2$ equal
to zero) and bond-bond ($\mbox{\boldmath $\delta$}_1$ and 
$\mbox{\boldmath $\delta$}_2$ different from zero) contributions to the
given {\bf q}-component of the Coulomb interaction. In eq.(\ref{a7}) we
have also passed to the adiabatic screened matrix elements, with the 
polarization diagrams given by 
\begin{eqnarray}
\label{a9}
\Pi_{l,l'}(\mbox{\boldmath $\delta$} - \mbox{\boldmath $\delta$}'; {\bf q},
\omega)= \frac{2}{N} \sum_{\bf k} e^{-i({\bf k} + \frac{{\bf q}}{2})
(\mbox{\boldmath  $\delta$}- \mbox{\boldmath $\delta$}')} 
\frac{1}{[1+S_l({\bf k})][1+S_{l'}({\bf k}+{\bf q})]} \times\nonumber \\
\times \frac{n_{l}({\bf k}) - n_{l'}({\bf k} + {\bf q})}{ \omega 
 +E_{l}({\bf k})- E_{l'}({\bf k}+{\bf q}) + i \eta}.
\end{eqnarray}

The coefficient on the left-hand sides of eq.(\ref{a7}),
\begin{equation}
\label{a10}
\varepsilon_{l_1l^{'}_1\mbox{\boldmath $\delta$}_1, l_3l^{'}_3
\mbox{\boldmath $\delta$}_3} \equiv 
\delta_{l'_{3}l_{1}} \delta_{l_{3}l^{'}_{1}}  
\delta_{\mbox{\boldmath $\delta$}_{3}\mbox{\boldmath $\delta$}_{1}} -
 \sum_{\mbox{\boldmath $\delta$}'_{3}} 
V_{l_{1}l_{3}l_{1}^{'}l'_{3}}(\mbox{\boldmath $\delta$}_1,
 \mbox{\boldmath $\delta$}^{'}_3; {\bf q})
\Pi_{l_{3}l_{3}^{'}}(\mbox{\boldmath $\delta$}^{'}_{3}-
\mbox{\boldmath $\delta$}_{3},{\bf q},\omega)
\end{equation}
define the dielectric matrix in the TB representation. Note that the sums
in eqs.(\ref{a7}) and (\ref{a10}) go over  
$\mbox{\boldmath $\delta$} = {\bf 0}$ and all nearest neighboring sites.
The order of the matrix (\ref{a10}) (i.e. the number of equations in the
system (\ref{a7}) for fixed $l_2$, $l^{'}_{2}$ and 
$\mbox{\boldmath $\delta$}_2)$ is equal to the number of pairs of band
indices $(l_1 l^{'}_{1})$ multiplied by $Z + 1$,where $Z$ is the number of
nearest neighbors (i.e. the coordination number for Bravais lattices).
Note that the polarization diagram (\ref{a9}) vanishes if $l$-th and
$l^{'}$-th bands are both full or empty, so that the number of relevant
pairs ($l l^{'}$) in eqs.(\ref{a8}) and (\ref{a10}) reduces to that
representing transitions between (partially) full and (partially) empty
bands (including the intra-band transtions).  Furthermore, the order of
the matrix (\ref{a10}) reduces to a finite number after a physical
truncation of the system (\ref{a7}) by which one keeps only contributions
from bands which are not too far energetically from the Fermi level, and
neglects e.g. those from deep core atomic states and very high empty bands. 

The determinant of the matrix (\ref{a10}) is of the central importance for
further considerations, since it carries  relevant information about the
microscopic dielectric properties. Indeed, it enters into the denominators
of all screened matrix elements, and, furthermore, its zeros define the
collective as well as the electron-hole excitations of the two-band 
electron gas \cite{zbbII}. For these reasons we call it microscopic
dielectric function, $\epsilon_{m}({\bf q},\omega) \equiv det[ \varepsilon]$. 

We summarize this Section by comparing the present TB formulation of the
dielectric matrix with the more usual one, based on the plane wave
representation and indexed by an infinite number of the vectors of
reciprocal lattice. The expressions (\ref{a7} - \ref{a10}) formally look
even less convenient, since the dielectric matrix (\ref{a10}) is spanned
by two multiplied infinities, the number of lattice sites (indices 
$\mbox{\boldmath $\delta$}$) and the number of intra- and inter-band
transitions [indices ($l l^{'}$)]. However, as argued above, both these
infinities are physically reduced to finite orders, which are eventually
rather low in standard TB systems.  No such reduction is possible
in the plane wave formulation, so that the order of the corresponding
dielectric matrix  remains physically very large. The reason is that the
base of reciprocal lattice vectors cannot be simply truncated whenever
the polarization processes take part on the scale of the unit cell.

\section { Two-band model}

The further analysis is concentrated on the case of two electron bands
which do not overlap. To fix ideas we assume that the lower ("valence")
band ($l = 0$) is partially or completely full, while the upper
("conducting") band ($l = 1$) is empty. The corresponding band dispersions
$E_l({\bf k})$ will be specified in the later examples \cite{zbbII}.
Furthermore, the later discussion will be mostly limited to the
long-wavelength regime (${\bf q} \rightarrow 0$). As it is shown in the
Appendix A, the contributions from the Coulomb matrix elements with bond
integrations [$\mbox{\boldmath $\delta$}_1$ and/or
$\mbox{\boldmath $\delta$}_2$ finite in eq.(\ref{a8})] are in this limit
weak and negligible in comparison to the site contributions. We therefore
continue by keeping only site matrix elements in the system (\ref{a7}),
and skip $\mbox{\boldmath $\delta$}$-indices from now on. The dielectric
matrix (\ref{a10}) then reduces to \cite{unna,giaq} 
\begin{equation}
\label{b1}
[ \varepsilon]= \left[ 
\begin {array}{cccc}
1-V_{0000}\Pi_{00} & -V_{0001}\Pi_{01}  & -V_{0100}\Pi_{10}  & 0 \\ 
-V_{0010}\Pi_{00}  & 1-V_{0011}\Pi_{01} & -V_{0110}\Pi_{10}  & 0 \\
-V_{1000}\Pi_{00}  & -V_{1001}\Pi_{01}  & 1-V_{1100}\Pi_{10} & 0 \\
-V_{1010}\Pi_{00}  & -V_{1011}\Pi_{01}  & -V_{1110}\Pi_{10}  & 1   
\end{array}
\right].
\end{equation}

The discussion of the dielectric matrix in previous works \cite{unna,giaq}
was restricted to the cases in which the off-diagonal matrix elements were
negligible. In the present work, we show that in the systems with finite
interband dipolar transitions these matrix elements are essential and can
by no means  be treated perturbatively.
On the other hand, the dielectric matrix discussed in Ref.\cite{unna}
allows for a Fermi surface crossing both bands due to their finite 
overlap. In our case of a finite gap between two bands, the matrix elements 
$\overline{V}_{1l_{2}1l_{2}^{'}}({\bf q},\omega)$ on the left-hand 
sides of (\ref{a7}) are simply multiplied by $\delta_{1l_1} 
\delta_{1l^{'}_1}$,  due to $\Pi_{11}({\bf q},\omega)=0$. 
The problem is therefore reduced to a system 
of three equations, as is seen from the form of the matrix (\ref{b1}).

A furher simplification in $[ \varepsilon]$ takes place after assuming
that the products of orbitals $\varphi^{*}_{l}({\bf r}) 
\varphi_{l^{'}}({\bf r})$  are real for each pair $(l, l^{'})$, as is
usually true in standard cases. Then the matrix elements of the 
screened interaction (\ref{a8}) obey symmetry relations 
\be
\label{b2}
\overline{V}_{l_{1}l_{2}l_{1}^{'}l_{2}^{'}}=
\overline{V}_{l_{1}^{'}l_{2}l_{1}l_{2}^{'}}=
\overline{V}_{l_{1}l_{2}^{'}l_{1}^{'}l_{2}}=\overline{V}_{l_{1}^{'}l_{2}^{'}l_{1}l_{2}},
\ee 
which are of course also fulfilled for the corresponding bare matrix
elements. Note also that 
\begin{equation}
\overline{V}_{l_{1}l_{2}l_{1}^{'}l_{2}^{'}}({\bf q},\omega)= 
\overline{V}_{l_{2}l_{1}l_{2}^{'}l_{1}^{'}}(-{\bf q},\omega),
\label{b3} 
\end{equation}
is generally valid due to $\overline{V}({\bf r},{\bf r'},\omega) = 
\overline{V}({\bf r'},{\bf r},\omega)$ and 
$\overline{V}({\bf r + R},{\bf r' + R},\omega)=\overline{V}({\bf r},
{\bf r'},\omega)$, where ${\bf R}$ is any lattice vector.

Due to the first  equality in (\ref{b2}), first three equations in  the
linear subsystem (\ref{b1}) reduce to two equations for 
$\overline{V}_{0l_{2}0l_{2}^{'}}$ and {\em e. g.}  
$\overline{V}_{0l_{2}1l_{2}^{'}}$. 
The corresponding matrix is given by
\begin{equation}
\label{b4}
\left[
\begin {array}{cccc}
1-V_{0000}\Pi_{00}  & -V_{0001}(\Pi_{01}+\Pi_{10}) \\
 -V_{1000}\Pi_{00}  & 1-V_{0011}(\Pi_{01}+\Pi_{10}) 
\end{array}
\right],
\end{equation}
while the microscopic dielectric function reads
\begin{eqnarray}
\epsilon_{m}({\bf q},\omega)&=&\left[1-V_{0000}({\bf q})
\Pi_{00}({\bf q},\omega)\right] \left\{1-V_{0011}({\bf q})
\left[ \Pi_{01}({\bf q},\omega)+\Pi_{10}({\bf q},\omega)\right]\right\} -
 \nonumber\\
 & &-V_{0001}({\bf q}) V_{1000}({\bf q})\Pi_{00}({\bf q},\omega)
\left[ \Pi_{01}({\bf q},\omega)+ \Pi_{10}({\bf q},\omega)\right].
\label{b5}
\end{eqnarray}

Finally, among sixteen TB matrix elements of the screened interaction
(\ref{a8}) six are given by following relations
\begin{eqnarray}
\overline{V}_{0000}&=&\frac{1}{\epsilon_{m}} \left[V_{0000}
+(V_{0001}V_{0010}- V_{0000}V_{0011})(\Pi_{01}+\Pi_{10})\right],
\label{b6}\\
\overline{V}_{0001}&=&\frac{1}{\epsilon_{m}}V_{0001},\label{b7}    \\
\overline{V}_{0011}&=&\frac{1}{\epsilon_{m}} \left[V_{0011}
+(V_{0001}V_{0010}-V_{0000} V_{0011})\Pi_{00} \right],\label{b8}\\
\overline{V}_{0101}&=&\frac{1}{\epsilon_{m}}\left[V_{0101}
+(V_{0111}V_{0001}- V_{0101}V_{0011})(\Pi_{01}+\Pi_{10})\right],\label{b9}\\
\overline{V}_{0111}&=&\frac{1}{\epsilon_{m}}\left[V_{0111}-
(V_{1010}V_{0001}- V_{1011}V_{0000})\Pi_{00}\right],\label{b10}\\
\overline{V}_{1111}&=&V_{1111}+\frac{1}{\epsilon_{m}} 
\left\{ V_{1010} \Pi_{00}          
\left[V_{0101}+(V_{0001}V_{0111}-V_{0101}V_{0011})(\Pi_{01}+\Pi_{10}) 
\right] \right. \nonumber\\
        & & \left. +V_{1110}(\Pi_{01}+\Pi_{10})
                   \left[V_{0111}+(V_{0010}V_{0101}-V_{0111}V_{0000})  
             \Pi_{00}\right] \right\},
\label{b11}
\end{eqnarray}
while the remaining ten matrix elements follow from the relations 
(\ref{b2}) and (\ref{b3}). 

The dimension of the  matrix (\ref{b4}) coincides with the number of 
intra-band and inter-band transitions for the present two-band model, 
in accordance with the general conclusion about the dimension of the
dielectric matrix in the tight-binding approach from Ref.\cite{zbb} and
Sect.2. Evidently, the result (\ref{b4}) simply reduces to the
one-dimensional matrices in the particular cases of the two-band
insulator (when $\Pi_{00} = 0$),
and of the one-band conductor (when  $\Pi_{01} = \Pi_{10} = 0$).

\section{Multipole expansion}

For further considerations it is appropriate to write the bare Coulomb matrix 
elements as sums of local and long-range (two-site)  parts,
\begin{equation}
V_{l_{1}l_{2}l_{1}^{'}l_{2}^{'}}({\bf q})=
U_{l_{1}l_{2}l_{1}^{'}l_{2}^{'}}+
W_{l_{1}l_{2}l_{1}^{'}l_{2}^{'}}({\bf q}),
\label{c1}
\end{equation}
with 
\be
\label{c2}
U_{l_{1}l_{2}l_{1}^{'}l_{2}^{'}}  = \int d{\bf r} \int d{\bf r'}
\varphi^{*}_{l_{1}}({\bf r})\varphi^{*}_{l_{2}}({\bf r'})
\frac{e^{2}}{\mid{\bf r}-{\bf r'}\mid}
\varphi_{l_{1'}}({\bf r})
\varphi_{l_{2'}}({\bf r'})
\ee
and 
\begin{equation}
W_{l_{1}l_{2}l_{1}^{'}l_{2}^{'}}({\bf q})= \sum_{\bf R \neq 0}
e^{i{\bf qR}}
\int d{\bf r}\int
 d{\bf r'}\varphi_{l_{1}}^{*}({\bf r}-{\bf R})\varphi_{l_{2}}^{*}({\bf r'})
\frac{e^{2}}{\mid{\bf r}-{\bf r'}\mid}
\varphi_{l_{1}^{'}}({\bf r}-{\bf R})
\varphi_{l_{2}^{'}}({\bf r'}).
\label{c3} 
\end{equation}
Integrations in the sums (\ref{c3}) are localized within the ranges of 
TB-orbitals. It was already noted that they are small not only with respect to 
the range of interaction, but also in comparison with the distances between 
neighboring crystal sites. It is therefore natural to perform the  
multipole expansions of the long-range parts. This is particularly useful
in the long wavelength limit in which the hierarchy of the leading multipole
terms is well defined.  In this Section we calculate dominant contributions
to the long-range bare matrix elements (\ref{c3}), and indicate for which
matrix elements (\ref{c1}) the on-site contributions (\ref{c2}) are relevant
for the dielectric response in the long wavelength limit.

Let us start from the expansion of the bare Coulomb interaction up to the
second order in the atomic coordinates \cite{jackson},
\begin{eqnarray}
\frac{1}
{\mid \mbox{\boldmath $\rho$}+{\bf R}-\mbox{\boldmath$\rho^{'}$}\mid}
&=&\frac{1}{R}-\frac{{\bf R}(\mbox{\boldmath $\rho$}
 -\mbox{\boldmath $\rho^{'}$})}{R^{3}}               
               +\frac{\mbox{\boldmath $\rho$} \mbox{\boldmath $\rho'$}-
     3(\mbox{\boldmath $\rho$} \cdot {\bf R_{0}})
(\mbox{\boldmath $\rho^{'}$} \cdot {\bf  R_{0}})}{R^{3}} \nonumber\\
& &-\frac{1}{2} \sum_{i,j}(\rho_{i}\rho_{j}+\rho_{i}^{'}\rho_{j}^{'})                      
\frac{R^{2}\delta_{i,j}-3X_{i}X_{j}}{R^{5} }+ ...,
\label{c4}
\end{eqnarray}
where ${\bf R}_{0}\equiv{\bf R}/R$, and {\boldmath $\rho$} and 
{\boldmath $ \rho^{'}$} are local electron positions at the {\bf R}-th
site and at the origin respectively.
$X_{i}$ and $\rho_{i}$ are i-th  Cartesian components of the vectors 
{\bf R} and {\boldmath $\rho$} respectively. In the next step we chose the 
parity of the TB orbitals by imposing
\be
\label{c5}
\varphi_{0}(\mbox{\bf -r})  = \varphi_{0}(\mbox{\bf r}),
\ee
and
\be
\label{c6}
\varphi_{1}(\mbox{\bf -r}) = \pm \varphi_{1}(\mbox{\bf r}).
\ee
Since the two possibilities defined by the upper [model A] and lower [model B]
sign in eq.(\ref{c6}) lead to qualitatively different properties of the
dielectric response, we shall treat both in parallel. Furthermore,
we choose the cubic lattice,  the simplest one for the calculation 
of lattice sums in the matrix elements of the terms in the expansion
(\ref{c4}). To fix ideas, we also specify in the further considerations
that $\varphi_{0}$ is an $s$ - orbital, while  $\varphi_{1}$ is an $d$
(upper sign) or $p$ (lower sign) orbital,
i.e. that the models A and B are characterized by $(s, d)$  and $(s, p)$
bands respectively. With these choices and   
with the assumption that there is no orbital degeneracy, the crystal
symmetry is strictly speaking lower than cubic. Still this is not a serious
inconsistency.
On the one side, one may suppose that the Bravais lattice is built from some
elongated molecules with e.g. $\sigma$ and $\pi$ electronic orbitals and that  
the differences in the lattice constants along three orthogonal crystal
directions are still small and have negligible effects on the lattice sums.
Alternatively, the present analysis can be straightforwardly completed by
including cubic orbital degeneracy into electronic spectrum. The dielectric
matrix for the cubic crystal with 
e. g. one  (partially) full $s$ and three empty $p$ bands will be derived
in the Appendix B. 

Let us now determine the leading terms in the expansion (\ref{c4}) of the
matrix elements (\ref{c1}), paying  particular attention to those present
in the dielectric matrix (\ref{b4}). 

$V_{0000}({\bf q})$ belongs to the  set of matrix elements 
$V_{l_{1}l_{2}l_{1}l_{2}}({\bf q})$ for scatterings in which 
electrons do not change bands. The leading long range contribution
$W_{l_{1}l_{2}l_{1} l_{2}}({\bf q})$ comes from the monopole-monopole
term which is same for all $l_{1}$ and $l_{2}$, no matter whether one
($l_{1} = l_{2}$) or two  ($l_{1} \neq l_{2}$) bands are involved. 
Since this term diverges in the limit ${\bf q} \rightarrow 0$, the on-site
contribution $U_{l_{1}l_{2}l_{1}l_{2}}$ to the total bare interaction
(\ref{c1}) can be neglected.
Keeping only the most divergent part of the monopole- monopole term, one gets  
for both models A and B
\be
V_{l_{1}l_{2}l_{1}l_{2}}({\bf q})=
 \frac{4\pi e^{2}}{a^{3}q^{2}},
\label{c7}
\ee
where $a^{3}$ is the volume of the primitive cell. 

The bare Coulomb matrix elements in which one or both electrons
change bands will depend on the choice of TB orbitals $\varphi_{0}$ and
$\varphi_{1}$. The decisive difference comes from the fact that the matrix
element for the dipole transition
\be                        
\mbox{\boldmath $\mu$}_{01}=\mbox{\boldmath $\mu$}^{*}_{10}
\equiv e\int d\mbox{\boldmath $\rho$}\,
\varphi^{*}_{0}(\mbox{\boldmath $\rho$})\mbox{\boldmath $\rho$}\varphi_{1}
(\mbox{\boldmath $\rho$})
\label{c8}
\ee
vanishes in the model A and is finite in the model B. We therefore 
continue by considering each model separately.

Let us start with the model B. The leading long range contribution to the
matrix elements in which one electron changes the band comes from the
monopole-dipole term in the expansion (\ref{c4}), and reads
\be
W_{lll l^{'}}({\bf q})=
  \sum_{{\bf R}\neq 0}e^{i{\bf qR}} 
\frac{\mbox{\boldmath $\mu$}_{ll^{'}} \cdot {\bf R}_{0}}{R^{2}}. 
\label{c9}
\ee
($l \neq l^{'}$). The divergent sum in this expression depends on the
angle between ${\bf q}$ and $\mbox{\boldmath $\mu$}_{ll^{'}}$, where the
direction of $\mbox{\boldmath $\mu$}_{ll^{'}}$ is determined by 
$\mbox{\bf r}$ - dependences in $\varphi_0$ and
$\varphi_1$ (\ref{c8}).  With $\varphi_1$ chosen to be the $p$ - orbital
elongated along the $x$ - axis,  $\mbox{\boldmath $\mu$}_{01}$ is also
directed along this axis. Noting in addition that in the model B the
on-site term $U_{llll^{'}}$ vanishes due to symmetry reasons, one gets 
\be
V_{llll^{'}}({\bf q})= - V_{l^{'}lll}({\bf q})=
\frac{4\pi ie \mu}{a^{3}} 
\frac{ q_{x}}{q^{2}}
\label{c10}
\ee
where $\mu \equiv |\mbox{\boldmath $\mu$}_{01}|$. The leading long 
range contribution  to matrix elements with both electrons changing bands 
is the dipole-dipole one, which after performing the dipolar
summations \cite{cohen} 
for the cubic crystal lattice in the limit ${\bf q} \rightarrow 0$, reads
\be
V_{lll^{'}l^{'}}({\bf q})=\frac{4\pi  \mu^{2}}{3a^{3}} 
\left(\frac{3q_{x}^{2}}{q^{2}}-1\right) + U_{lll^{'}l^{'}},
\label{c11}
\ee
with $l \neq l^{'}$. Eqs.(\ref{c7}), (\ref{c10}) and (\ref{c11}) exhaust
all matrix elements for the model B. We conclude that for 
${\bf q} \rightarrow 0$
the on-site and the long range contributions enter on the equal footing
only into the matrix elements (\ref{c11}), while in the matrix elements
(\ref{c7}) and (\ref{c10})
the former are negligible with respect to the latter.

In the model A the leading long range contributions to the matrix elements
$V_{llll^{'}}({\bf q})$ 
and $V_{lll^{'}l^{'}}({\bf q})$ come from  monopole-quadrupole and
quadrupole-quadrupole terms in the expansion (\ref{c4}), respectively.
If the matrix elements of the quadrupolar transition between the orbitals
$\varphi_{l}$ and  $\varphi_{l'}$ are finite, these contributions behave
in the limit ${\bf q} \rightarrow 0$ as $q^{0}$ and $q^{2}$, respectively.
We do not need their detailed forms, since the further conclusions will be
based solely on the fact the corresponding total matrix elements 
$V_{llll^{'}}({\bf q})$ and $V_{lll^{'}l^{'}}({\bf q})$ are regular
in the long wavelength limit (and $V_{lll^{'}l^{'}} \simeq U_{lll^{'}l^{'}}$
in addition).

The most important difference between the $\bf q \rightarrow 0$ limits of
models A and B appears in  processes in which one electron changes its
band. While in the model B
the corresponding matrix element (\ref{c10}) comes from the long range part and 
diverges as $q^{-1}$, the analogous matrix element in the model A is the
sum of the long range part and the on-site contribution and behaves as $q^{0}$. 

It remains to determine the asymptotic behavior of the polarization diagrams
which figure in the dielectric matrix (\ref{b4}). We limit the present
discussion to the regime (${\bf q} \rightarrow 0, \omega \neq 0$) in which
the real parts of the polarization diagrams are given by the well-known
expressions
\begin{equation}
  \mbox{Re} \; \Pi_{00}({\bf q},\omega)=\frac{n_{e}}{m^{*}} 
\frac{q^{2}}{ \omega^{2}},
\label{c12} 
\end{equation}
and
\begin{equation}
  \mbox{Re} [\Pi_{01}({\bf q},\omega)+\Pi_{10}({\bf q},\omega)]=2 n_{e} 
 \frac{E_{g}}{\omega^{2}-E_{g}^{2}}.
 \label{c13}
\end{equation}
$m^{*}$ is the effective mass and $n_{e}$ is the number of electrons per 
site in the lower band, while $E_{g} \equiv E_{1} - E_{0}$ is the energy
difference between the band centers. Here we neglect the bandwidths in the
expression (\ref{c13}) assuming that they are  much smaller than $E_{g} $.
Note that the expressions (\ref{c12}) and (\ref{c13}) do not 
depend on the symmetry of TB orbitals $\varphi_{0}$ and $\varphi_{1}$.
Furthermore, we do not introduce the corresponding imaginary parts,
relegating the discussion of the effects of Landau damping to the paper II.
We only note that in the above $({\bf q},\omega)$-regime Im $\Pi_{00} = 0$
while Im$[\Pi_{01} + \Pi_{10}]$ may be finite. 

We are now ready to compare the microscopic dielectric functions for models
A and B. Let us start with the simpler case of model A, already considered 
in Ref.(\cite {unna}) and explored in Refs.(\cite {imry1,imry2}).
As it is seen from (\ref{c12}) and (\ref{c13}), the term in the
expression (\ref{b5}) which comes from the 
off-diagonal elements in the matrix (\ref{b4}) vanishes in the long
wavelength limit.  In other words, the intra-band and inter-band
polarization processes, given by the
diagonal elements in this matrix, become decoupled, and it is legitimate to 
distinguish between the intra-band ($\epsilon_{intra}$) and inter-band
($\epsilon_{inter}$) dielectric functions \cite {unna}, the total
dielectric function being the product of the two.
Due to this factorization the collective modes, defined by
the real zeros of $\epsilon_{m}$ in the $\omega$-plane, are also either of
intraband or of interband origin. They will be analysed in detail in the
paper II. Furthermore, the expressions (\ref{b6}-\ref{b11}) for the
screened Coulomb interaction reduce to 
\begin{equation}
\overline{V}_{ll'll'}/V_{ll'll'} = 
 1/ \epsilon_{intra}
\label{c14}
\end{equation}
(with either $l = l'$ or $l \neq l'$), and
\begin{equation}
\overline{V}_{llll'}/V_{llll'} =
\overline{V}_{lll'l'}/V_{lll'l'} =  1/ \epsilon_{inter}
\label{c15}
\end{equation}
(with $l \neq l'$). Thus all purely intraband scatterings are screened by 
$\epsilon_{intra}$, while the scatterings with at least one interband
electron transition are screened by $\epsilon_{inter}$.

The above decoupling does not hold in the model B.
The product of two off-diagonal matrix elements in (\ref{b4}) now remains
finite in the long wavelength limit, and introduces a coupling between the
intraband and interband contributions into the microscopic dielectric
function (\ref{b5}). The present analysis shows that this coupling comes
from the divergent monopole-dipole contribution (\ref{c10}) to the matrix
element $V_{llll'}({\bf q})$.
Its most important consequence, the hybridization of the intraband and 
interband collective modes in conductors, will be considered in the paper II. 

\section{Macroscopic dielectric function}

The aim of this section is to establish the relation between the
macroscopic dielectric function $\epsilon_{M}({\bf q},\omega)$ and our
microscopic dielectric function  
$\epsilon_{m}({\bf q},\omega)$. The former is defined in a standard way as 
the ratio between the Fourier components of the spatial averages of the bare 
(external) and screened (total) Coulomb 
potential. To get macroscopic  averages, it is enough to take the unit cell
as the averaging volume \cite{wiser}. Let us denote the averaged bare and
screened interactions between the unit cells at sites ${\bf R}$ and 
${\bf R'}$ by ${V}_{av}({\bf R,R'})$ and 
$\overline{V}_{av}({\bf R,R'},\omega)$ respectively. Since at long
wavelengths details in the space dependence of the true (non-averaged)
interactions on the scale of unit cell  are irrelevant, the true and averaged 
interactions are in this limit indistinguishable, and one gets the connection
\begin{equation}
\overline{V}_{av}({\bf q +G, q + G'},\omega) \cong 
\delta_{{\bf G},0} \delta_{{\bf G'},0}\overline{V}({\bf q,q},\omega)
\label{d1}
\end{equation}
between the corresponding Fourier transforms. Here ${\bf G}$ and ${\bf G'}$
denote the vectors of the reciprocal lattice. The same relation holds for
the averaged bare interaction $V_{av}({\bf q})$. 

The macroscopic dielectric function  is now defined by
\begin{equation}
\label{d2}
\frac{1}{\epsilon_{M}({\bf q},\omega)}=
\frac{\overline{V}_{av}({\bf q},{\bf q},\omega)}{V_{av}({\bf q})} = 
\epsilon^{-1}({\bf q,q},\omega).
\end{equation}
The ratio $\overline{V}_{av}({\bf q},{\bf q},\omega)/V_{av}({\bf q})$ can
be also interpreted as the ${\bf q}$-component of the total charge density, 
redistributed due to the Coulomb screening,  averaged across the unit cell and 
divided by the corresponding ${\bf q}$-component of the probe
charge \cite{zbb}. 
The last equality in eq.(\ref{d2}) follows from the standard definition
of the dielectric matrix in the plane wave representation,
$\overline{V}({\bf q,G,G'},\omega)=\epsilon^{-1}({\bf q+G',q+G},\omega)V
({\bf q,G,G})$,
and the relation (\ref{d1}). Our aim is however to express
$\epsilon_{M}({\bf q},\omega)$ in terms of the TB bare and screened
Coulomb matrix elements (\ref{c1}, \ref{a8}). The TB matrix elements
(\ref{a8}) can be expressed as double sums over the Fourier transforms 
$\overline{V}({\bf q + G},{\bf q + G'},\omega)$ through the relation
\begin{equation}
\label{d3}
\overline{V}_{l_{1}l_{2}l_{1}'l_{2}'}({\bf q},\omega)= \sum_{{\bf G},{\bf G'}} 
I_{l_{1}l_{1}'}({\bf q+G})I_{l'_{2}l_{2}}^{*}({\bf q+G'})
\overline{V}({\bf q + G},{\bf q + G'},\omega)
\end{equation}
with 
\begin{equation}
\label{d4}
I_{ll'}({\bf q}) \equiv \int d{\bf r} \,
\varphi^{*}_{l}({\bf r})   e^{i{\bf qr}}  \varphi_{l'}({\bf r}).
\end{equation}
The corresponding relation for the averaged matrix elements follows after
inserting the relation (\ref{d1}) into (\ref{d3}). Expanding the matrix
elements (\ref{d4}) in powers of ${\bf q}$ we get in the limit 
${\bf q} \rightarrow 0$
\begin{equation}
\overline{V}_{av,l_{1}l_{2}l_{1}^{'}l_{2}^{'}}({\bf q},\omega ) \cong 
\left [ \delta_{l_{1},l_{1}^{'}}+\frac{i{\bf q} 
\mbox{\boldmath $\mu$}_{l_{1}l_{1}^{'}}}{e} + {\em O}({\bf q}^{2})\right ]
\left [ \delta_{l_{2},l_{2}^{'}}-\frac{i{\bf q} 
\mbox{\boldmath $\mu$}_{l_{2}l_{2}^{'}}}{e} + {\em O}({\bf q}^{2})\right ]
\overline{V}_{av}({\bf q},\omega)
\label{d5}
\end{equation}
with $\mbox{\boldmath $\mu$}_{l_{1}l_{1}^{'}}$ being the dipolar matrix
element (\ref{c8}). Thus, whenever 
$l_{1} \neq l_{1}^{'}$ and/or $l_{2} \neq l_{2}^{'}$, 
the corresponding ratios 
$\overline{V}_{av,l_{1}l_{2}l_{1}^{'}l_{2}^{'}}({\bf q},\omega)/
\overline{V}_{av}({\bf q},\omega)$ tend to zero as ${\bf q} \rightarrow 0$.
In other words, since the averaged interactions do not vary within a unit
cell, the contributions from the
dipolar and higher transitions  to the corresponding TB matrix elements 
$\overline{V}_{av,l_{1}l_{2}l_{1}^{'}l_{2}^{'}}({\bf q},\omega)$ vanish.
For $l_{1}=l_{1}^{'}$ and $l_{2}=l_{2}^{'}$, i.e. for matrix elements
with a finite monopole-monopole contribution, we have
\begin{equation}
\label{d6}
\overline{V}_{av,l_{1}l_{2}l_{1}l_{2}}({\bf q},\omega )
\cong \overline{V}_{l_{1}l_{2}l_{1}l_{2}}({\bf q},\omega ) 
\cong\overline{V}_{av}({\bf q},\omega)
\end{equation}
for any pair of orbital indices $l_{1}$ and $l_{2}$. The first equality in
(\ref{d6}) is based on the same argument as the relation (\ref{d1}). Since
the relations (\ref{d5}) and (\ref{d6}) are also valid for the bare
interaction $V({\bf r}-{\bf r^{'}})$, the macroscopic dielectric function
reads
\begin{equation}
\label{d7}
\epsilon_{M}({\bf q},\omega)=
\frac{V_{l_{1}l_{2}l_{1}l_{2}}({\bf q})}
{\overline{V}_{l_{1}l_{2}l_{1}l_{2}}({\bf q},\omega )}.
\end{equation}
This result is generally valid for any number of bands.  It also does not 
depend on the method of calculation of the linear dielectric response,
 which brings in a particular relationship between the screened and bare 
Coulomb interaction. The independence of the right-hand side of (\ref{d7})
on the indices $l_{1}$ and $l_{2}$ may serve as a check of the consistency of 
a given approximation in the macroscopic limit. It is easy to see that this
independence is  realized for the RPA result (\ref{b6}-\ref{b11}).

Although $\epsilon_{M}({\bf q},\omega)$ is the ratio of the bare and screened 
monopole-monopole interactions, this of course does not mean that other
higher order terms in the multipole expansion do not contribute to the
macroscopic response. This becomes clear already for the simple two-band
models from Sects. 3, 4. For the model A the result (\ref{c14}) reads
\begin{equation}
\label{d8}
\epsilon_{M}({\bf q},\omega)=\epsilon_{intra}({\bf q},\omega)
= 1+4 \pi \alpha_{c}({\bf q}, \omega),
\end{equation}
with 
\begin{equation}
\label{d9}
\alpha_{c}({\bf q}, \omega) \equiv -\frac{e^{2}}{4 \pi q^{2}} 
\Pi_{00}({\bf q},\omega)
\end{equation}
representing the intraband polarizability. Thus, the macroscopic dielectric 
function is entirely determined by the intraband processes, and reduces to
unity when the lower band is fully occupied. In other words, the interband
modes are not directly optically active, and can possibly be observed only
by Raman scattering.

The macroscopic dielectric response  for the model B follows after 
inserting any of the expressions (\ref{b6},\ref{b9},\ref{b11}) into
(\ref{d7}), and taking into account the long wavelength asymptotic 
expressions (\ref{c7},\ref{c10},\ref{c11}). One gets 
\begin{equation}
\label{d10}
\frac{\epsilon_{m}} {\epsilon_{M}}= 1 +  
\left [\frac{4\pi\mu^{2}}{3a^{3}} - U_{0011}\right ] (\Pi_{01}+\Pi_{10}).
\end{equation}
Generally, the right-hand side in this expression depends on the direction
of ${\bf q}$ due to the anisotropy of the polarization diagram 
$(\Pi_{01}+\Pi_{10})$. However, in the particular limit given by
eq.(\ref{c13}) this is not the case, so that the orientational dependence
of $\epsilon_{M}({\bf q }, \omega)$ is determined solely
by that of the microscopic function $\epsilon_{m}({\bf q } , \omega)$. 
Let us consider here only two particular directions, 
${\bf q} \perp  \mbox{\boldmath $\mu$}$
and ${\bf q} \parallel \mbox{\boldmath $\mu$}$.

For ${\bf q}={\bf q}_{\perp} \perp \mbox{\boldmath $\mu$}$ the bare
monopole-dipole matrix element (\ref{c10}) vanishes, so that
$\epsilon_{m} = \epsilon_{intra}\epsilon_{inter}$
with $\epsilon_{inter}$ equal to the right-hand side od (\ref{d10}).
Hence one gets
\begin{equation}
\label{d11}
\epsilon_{M}({\bf q}_{\perp}, \omega) = 
1+4 \pi \alpha_{c}({\bf q}_{\perp},\omega),
\end{equation}
as should be expected, since for ${\bf q} \perp \mbox{\boldmath $\mu$}$  
the interband  longitudinal  polarization processes  are not possible.

For ${\bf q} \parallel \mbox{\boldmath $\mu$}$ the (intraband)
monopole - (interband) dipole coupling is finite and, in addition, even
for an insulator there is no simplification
in (\ref{d10}) as for ${\bf q} \perp \mbox{\boldmath $\mu$}$. The macroscopic 
dielectric function can be then written in the form
\begin{equation}
\label{d12}
\epsilon_{M}(q_{\parallel},\omega)=  1+4 \pi \alpha_{c} +
\frac{ 4 \pi \overline{\alpha_{I}} }{1- \frac{4 \pi}{3} 
\overline{\alpha_{I}} }.
\end{equation}
Here 
\begin{equation}
\label{d13}
\overline{\alpha_{I}}( q_{\parallel},\omega) \equiv
 \frac{\alpha_{I}( q_{\parallel},\omega)}
{1 + U_{0011}a^{3}\alpha_{I}( q_{\parallel},\omega) /\mu^{2}}, 
\end{equation}
with ${\alpha_{I}( q_{\parallel},\omega)}$ being the dipolar interband
polarizability
\begin{equation}
\label{d14}
\alpha_{I}( q_{\parallel},\omega)\equiv -\frac{\mu^{2}}{a^{3}} 
\left [ \Pi_{01}( q_{\parallel},\omega)
 +\Pi_{10}( q_{\parallel},\omega \right)].
\end{equation}

The three quantities which enter into the expressions (\ref{d13}, \ref{d14})
represent the intraband ($\alpha_{c}$), interband ($\alpha_{I}$) and 
intra-atomic [$U_{0011}(\Pi_{01}+\Pi_{10})$]
contributions to the macroscopic dielectric response.  Consequently, 
this expression covers, and interpolates between, limiting cases  of
one-band conductor,  two (multi)-band insulator, and atomic (zero bandwidth)
insulator.
Before going into particular limits, we remind that Adler \cite{adler} 
parametrized  any deviation of the expression (\ref{d12}) from the usual
form containing only the dipolar interband polarizability (\ref{d14}) by
introducing a self-polarization correction, defined as 
$\alpha_{I}( q_{\parallel},\omega) - \overline{\alpha_{I}}( q_{\parallel},
\omega)\equiv C$.
When expressed in terms of TB quantities, this parameter reads
\begin{equation}
\label{d15}
C=\frac{\mu^{2}}{a^{3}}
\frac{U_{0011}(\Pi_{01}+\Pi_{10})^{2}}{1-U_{0011}(\Pi_{01}+\Pi_{10})}.										
\end{equation}
As it will be shown below, the TB results (\ref{d13}) and (\ref{d14}) provide
a direct physical insight into this correction for atomic insulators as
well as for band insulators and conductors.

Let us now consider some characteristic limits. 
The usual Lindhard (i.e. Sellmeyer \cite{dar})  expression for a metal
follows after
neglecting interband polarization ($\alpha_{I} = 0$), while the opposite
limit for an
atomic insulator is obtained after putting $\alpha_{c} = 0$ and neglecting all 
bandwidths \cite{giaq}. The latter result is the well known Lorentz-Lorenz
(i.e. Clausius-Mossotti) expression. In this limit the polarizability 
$\overline{\alpha_{I}}$ which enters into eq.(\ref{d12}) is given by
\begin{equation}
\label{d16}
\overline{\alpha_{I}}({\bf q},\omega) = 
 -\frac{2n_{e}}{a^{3}}\frac{E_{g,eff}\mu_{eff}^{2}}{\omega^{2} - 
E_{g,eff}^{2}},
\end{equation}
i.e. it does not differ in form from the expression (\ref{d14},\ref{c13}). 
Here $\mu_{eff}^{2} = \mu^{2}E_{g}/E_{g,eff}$ and  $E_{g,eff}^{2} = E_{g}^{2}
+ 2nE_{g}U_{0011}$ are the renormalized values of the dipole matrix element
and the level spacing, i.e. instead of the initial (e.g. atomic
Hartree-Fock) parameters one obtains those which 
include the on-site RPA screening. In other words, the Adler's
self-polarization correction in this case simply reduces to a
renormalization of the parameters in the  dipolar (intra-atomic)
polarizability. This reflects the fact that the RPA scheme proposed 
in the present work treats both nonlocal and on-site screenings at the
same level of approximation. In this respect, we warn that in the limit
 (\ref{d16}) the RPA is usually not a reliable scheme at the atomic
(or molecular) level \cite{cederbaum}, and that it is obviously invalid
when the on-site interactions are large in comparison with $E_{g}$.

In the cases of insulators and conductors with finite bandwidths the
effective dipolar polarizability (\ref{d13}) does not have  the simple
form (\ref{d16}) and, moreover, cannot be reduced to the form of an
effective interband polarization diagram, cf. (\ref{d14}). Instead, one
has a nontrivial $\omega$-dependence in the denominator of the
expression (\ref{d13}). The self-polarization correction is no more of
entirely local (i.e. intra-atomic) origin, but appears to be a combined
effect of the on-site Coulomb interaction $U_{0011}$ in which two
electrons exchange their orbital states, 
and of the finite inter-site tunneling (i.e. of the finite bandwidths).  
The role of $U_{0011}$ in the macroscopic dielectric function was already 
noticed in Ref.\cite{giaq}. We note 
that some later results for $\epsilon_{M}$ \cite{sinha,onodera}, although
based on the RPA method,  cannot be expressed in the form 
(\ref{d13}, \ref{d14}) i.e. (\ref{d15}). The reason might be traced in the
additional approximations  which, unlike  the present RPA scheme, do not 
include the local on-site screening on the same footing with the long range
one.

Furthermore, it is appropriate to point out that $\overline{\alpha_{I}}$
appears in the denominator of eq.(\ref{d12}) due to the finiteness of
${\bf G}\neq0$ terms (local fields)
in the Fourier transform of the dipole-dipole interaction 
(\ref{c11}). The ${\bf G}\neq0$ terms are comparable to the 
\mbox{${\bf G}=0$} contribution
and should not be neglected, as already stressed in the Introduction.
Their omission transforms the formula (\ref{d12})  into an expression which
is additive in the polarizabilities \cite{nozieres,pines}.

\section{Conclusion}

The most important step in the present work is the use of the complete set
of TB states in the representation of the bare and screened Coulomb
interactions.
In this basis, the symmetry properties of the Coulomb matrix elements
follow directly from the symmetry of molecular orbitals participating
in the band states.  As a consequence,  the initial system of linear equations
for the screened matrix elements decomposes into smaller subsets 
with a common  dielectric matrix. E.g., in the two-band case sixteen linear
equations were reduced to two equations. This agrees with our general
conclusion \cite{zbb} that for a multi-band system with only two-center
Coulomb interactions taken into account, 
the TB dielectric matrix within RPA has a dimension equal to the
number of non-vanishing intra-band and inter-band polarization
diagrams (\ref{a9}).

The present approach is particularly convenient in the long wavelength limit, 
in which the multipole expansion is feasible. Again, the symmetry of
molecular orbitals is of central importance, since it determines the degree
of mixing between intra-band and inter-band polarization processes in
multiband conductors. In particular, whenever the interband processes are
dipolar the off-diagonal elements in
the TB dielectric matrix are finite and cannot be treated perturbatively.
The inclusion of  the long-range monopole-dipole interaction induced by
these processes leads to the correct expression for the macroscopic
dielectric function, and, as will be shown in paper II \cite{zbbII},
essentially influences the spectrum of collective modes 
for a multiband electron liquid (see also Ref. \cite{zbb}). In this respect
it is important to recognize that the results for 
$\epsilon_m({\bf q}, \omega)$ (\ref{a10}) and 
$\epsilon_M({\bf q}, \omega)$ (\ref{d12}) reduce to the additive 
form \cite{nozieres,pines}
\begin{equation}
\epsilon({\bf q},\omega)-1=4\pi\alpha_{c}+4\pi \sum_{I}\alpha_{I},
\label{con1}
\end{equation}
only in the high frequency limit, i.e. for frequencies much higher than
all interband energy differences $E_{gI}$. [The $I$-summation in
eq.(\ref{con1}) goes over all relevant interband polarizabilities
(\ref{d14})]. It is interesting to note that in this limit we apparently
recover the well-known expression for the high frequency plasma edge
expressed through the standard plasmon sum rule, involving the free
electron mass.

In conclusion, we point out that the present approach might have numerous
applications, particularly in systems with dipolar degrees of freedom 
which are sensitive to the variations of the valence band properties 
induced by pressure, changes in band filling, etc. We expect  that the 
present method  will prove  very useful in such investigations.

\appendix
\section{Appendix A}
\setcounter{equation}{0}
\renewcommand{\theequation}{A.\arabic{equation}}

In this Appendix we consider the influence of the site-bond
({\boldmath $\delta_1$} or {\boldmath $\delta_2$} different from zero)
and bond-bond  ({\boldmath $\delta_1$} and {\boldmath $\delta_2$}
different from zero) Coulomb interaction (\ref{a8}) on the dielectric
response in the long-wavelength limit (${\bf q} \rightarrow 0$,
 $\omega$ finite).  We take the simplest examples of one-band and
 two-band systems with the one-dimensional lattice
[{\boldmath $\delta_i$} in eqs.(\ref{a7} - \ref{a10}) equal to 0
 and $\pm1$]. The 
extensions to more complex band structures and lattices are straightforward. 

The system (\ref{a7}) simplifies after assuming, like in Sects.3 and 4,
 that the orbitals
$\varphi_l({\bf r})$ are even or odd, and that the products 
$\varphi_{l_{1}}^{*}({\bf r}-{\bf \delta})\varphi_{l_{2}}({\bf r})$
 are real. This leads
to additional symmetry relations between bare Coulomb matrix elements
 (\ref{a8}), which in the case of two-band system ($l_i= 0, 1$) read
\begin{eqnarray}
\label{app1}
V_{l_1l_2l'_1l'_2}(\mbox{\boldmath $\delta$}_1,\mbox{\boldmath $\delta$}_2) =
 (-1)^{\sum l_i} V_{l_2l_1l'_2l'_1}(\mbox{\boldmath $\delta$}_2, 
\mbox{\boldmath $\delta$}_1)
= (-1)^{\sum l_i} V_{l'_2l'_1l_2l_1}(-\mbox{\boldmath $\delta$}_2, 
-\mbox{\boldmath $\delta$}_1) = \nonumber\\
= V_{l_1l'_2l'_1l_2}(\mbox{\boldmath $\delta$}_1, -\mbox{\boldmath $\delta$}_2)
= V_{l'_1l_2l_1l'_2}(-\mbox{\boldmath $\delta$}_1, 
\mbox{\boldmath $\delta$}_2) =
 V_{l'_1l'_2l_1l_2}(-\mbox{\boldmath $\delta$}_1, 
-\mbox{\boldmath $\delta$}_2). 
\end{eqnarray}
 Due to these symmetry relations, the 
number of equations in (\ref{a7})
 reduces to $N_e/2$ or $(N_e+1)/2$ for even or odd  $N_e$ 
respectively, where  
 $N_e$ is the product of $ Z+1$ and the number of the non-zero 
 polarization diagrams among $\Pi_{00}, \Pi_{01}$ and $ \Pi_{10}$.
 
In particular,  taking into account these relations, we obtain the
microscopic dielectric function for 
one-band metal in the form of a $2\times 2$ determinant
\begin{equation}
\label{app2}
\varepsilon_{m}=
\left| \begin{array} {ll}
1-\sum_{\mbox{\boldmath $\delta$}}V(0,\mbox{\boldmath $\delta$}) 
\Pi_{0}(\mbox{\boldmath $\delta$}) & -\sum_{\mbox{\boldmath $\delta$}}
V(1,\mbox{\boldmath $\delta$}) \Pi_{0}(\mbox{\boldmath $\delta$}) \\
 -\sum_{\mbox{\boldmath $\delta$}}V(0,\mbox{\boldmath $\delta$})
\left[ \Pi_{0}(\mbox{\boldmath $\delta$}-1)+\Pi_{0}(\mbox{\boldmath 
$\delta$}+1) \right] & 
1-\sum_{\mbox{\boldmath $\delta$}}V(1,\mbox{\boldmath $\delta$})
\left[ \Pi_{0}(\mbox{\boldmath $\delta$}-1)+\Pi_{0}(\mbox{\boldmath 
$\delta$}+1) \right] 
\end{array}
\right|
\end{equation}
with $V(\mbox{\boldmath $\delta$}_1, \mbox{\boldmath $\delta$}_2)
 \equiv V_{0000}(\mbox{\boldmath $\delta$}_1, \mbox{\boldmath $\delta$}_2,
{\bf q})$ and 
$\Pi_{0}(\mbox{\boldmath $\delta$})\equiv \Pi_{00}(\mbox{\boldmath
 $\delta$}, {\bf q}, \omega)$. In the limit ${\bf q} \rightarrow 0$
the leading monopole-monopole terms in the bare matrix elements appearing
 in (\ref{app2}) are simply related by
\begin{equation}
\label{app3}
V(1,1,{\bf q}) = S V(1,0,{\bf q}) = S^2 V(0,0,{\bf q})  = S^2  
\frac{4\pi e^2}{a^3 q^2},
\end{equation}
where $S$ is the direct overlap (\ref{a6}). After inserting these relations
 and the expressions
(\ref{a9}) for the polarizability diagrams $\Pi_{0}(0), \Pi_{0}(\pm 1)$ and
 $\Pi_{0}(\pm 2)$ into the expression (\ref{app2}), it reduces to 
\begin{equation}
\label{app4}
\varepsilon_{m}= 1 - V(0,0,{\bf q}) \Pi_{0}(0)  \left[ 1 + {\em O}(q^2) \right].
\end{equation}
Thus, we conclude that in the limit ${\bf q} \rightarrow 0$, $\omega$
 finite, the microscopic dielectric
function for a one-band metal does not depend on the overlap factor $S$
 which enters into the 
site-bond and bond-bond interactions through the expressions (\ref{app3}).
In other words, it is completely determined by the site-site correlations.

In the case of the two-band insulator the most important simplification
 comes from the fact that
the only non-vanishing polarization diagrams are the inter-band ones with
 the zero phase factor
[$\mbox{\boldmath $\delta$} - \mbox{\boldmath $\delta$}' ={\bf 0}$ in
 eq.(\ref{a9})].  The microscopic dielectric function then reduces to 
\begin{equation}
\label{app5}
\varepsilon_{m}= 
\left|
\begin{array}{lll}
1-V_{1}(1,1)\Pi_{1} &-V_{1}(0,1)\Pi_{1} & -V_{1}(-1,1)\Pi_{1}\\
 -V_{1}^{*}(0,1)\Pi_{1} &1-V_{1}(0,0)\Pi_{1} & -V_{1}(0,1)\Pi_{1}\\
 -V_{1}^{*}(-1,1)\Pi_{1} &-V_{1}^{*}(0,1)\Pi_{1} &1 -V_{1}(1,1)\Pi_{1}
\end{array}
\right|
\end{equation}
with $V_{1}(\mbox{\boldmath $\delta$}, \mbox{\boldmath $\delta$}')
 \equiv V_{1001}(\mbox{\boldmath $\delta$}, \mbox{\boldmath $\delta$}';
 {\bf q})$. Since only one polarization factor
[$ \Pi_{01}(0,{\bf q},\omega) +  \Pi_{10}(0,{\bf q},\omega)\equiv  
\Pi_{1}$] figures in 
this determinant,  the criterion of weakness of site-bond and bond-bond
 correlations with
respect to the site-site ones obviously involves only the corresponding
 matrix elements of bare 
Coulomb interaction, i.e. $V_{1}(0,1)$, $V_{1}(1,1)$ and $V_{1}(-1,1)$ 
{\em vs} $V_{1}(0,0)$.
More precisely, the three- and four-center correlations may be considered
as weak perturbations provided the matrix element of bond dipolar transition 
\begin{equation}
\label{app6}
\mu(\mbox{\boldmath $\delta$}) =
 \int d{\bf r} \varphi^{*}_{0}( {\bf r}) {\bf r} 
\varphi_{1}({\bf r}-\mbox{\boldmath $\delta$})
\end{equation}
and local site-bond and bond-bond matrix elements are much smaller 
than the respective site parameter (\ref{c8}) and $U_{0011}$ 
[see eq.(\ref{c11})].

For a two-band metal $\varepsilon_m$ is given by a 5$\times$5 determinant, 
not written here for the sake of space. Although this determinant contains
intraband and interband polarization diagrams (\ref{a9}) with various phase
factors, the analysis of
the long wavelength limit leads to the same conclusion, namely that 
the criterion for the weakness of site-bond and bond-bond correlations in the 
dielectric function involves only the corresponding bare Coulomb matrix
elements.

\section{Appendix B}

\setcounter{equation}{0}
\renewcommand{\theequation}{B.\arabic{equation}}

In Sections 4 and 5 we have assumed within the model B that there is 
one orbital with $p_{x}$ symmetry at each site of a 
cubic Bravais lattice. In order to complete the discussion,
we take here the simplest generalization by which the
electron bands have the cubic symmetry.  We assume that there are  
one lower $s$-orbital (\ref{c5}) and three degenerate upper $p$ - orbitals
with the symmetry 
\begin{equation}
\label{ap1}
\varphi_{x}(-x,y,z) = - \varphi_{x}(x,y,z)\\
\end{equation}
and equivalently for $x \rightarrow y \rightarrow z$.

Following the procedure from Sec.3 we get for this four-band model
a linear Dyson system (\ref{a7}) which contains 256 equations and is 
decomposed into 16 subsystems of 16 equations. 
After assuming,  like in Sec.3, that all products 
 $\varphi^{*}_{l_{i}}({\bf r}) \varphi_{l^{'}_{i}}({\bf r})$ are real, 
the problem reduces to a system of four equations 
\begin{eqnarray}
\overline{V}_{000x}&=&V_{000x}+V_{0000}
\overline{V}_{000x}\Pi_{00}+V_{000x} \overline{V}_{00xx}(\Pi_{0x}+\Pi_{x0}),
\label{ap2}\\
 \overline{V}_{00xx}&=&V_{00xx}+V_{x000} \overline{V}_{000x}\Pi_{00}
+V_{00xx} \overline{V}_{00xx}(\Pi_{0x}+\Pi_{x0}), 
\label{ap3}\\
\label{ap4}
 \overline{V}_{00yx}&=&V_{00yy} \overline{V}_{00yx}(\Pi_{0y}
+\Pi_{y0}), \\
 \overline{V}_{00zx}&=&V_{00zz} \overline{V}_{00zx}
(\Pi_{0z}+\Pi_{z0}). 
\label{ap5}
\end{eqnarray}
Here subscripts $x,y$ and $z$ stand for $p_{x},p_{y}$ and $p_{z}$ orbitals
respectively, and the wave vector {\bf q} is oriented along the
crystal axis $x$. Due to the isotropy of the cubic crystal
this special choice does not harm the generality of the system
(\ref{ap2}-\ref{ap5})
(see also Ref.\cite{zbb}).

The corresponding dielectric matrix is given by 
\normalsize
\begin{equation}
\label{ap6}
[\varepsilon_{m}]= \left[ 
\begin {array}{cccc}
1-V_{0000}\Pi_{00} & -V_{000x}(\Pi_{0x}+\Pi_{x0}) & 0 & 0  \\
 -V_{x000}\Pi_{00} & 1-V_{00xx}(\Pi_{0x}+\Pi_{x0}) & 0 &0 \\  
0                  & 0 & 1-V_{00yy}(\Pi_{0y}+\Pi_{1y0}) & 0 \\
0                  & 0 & 0 & 1-V_{00zz}(\Pi_{0z}+\Pi_{z0})
\end{array}
\right],
\end{equation}
\large
and all TB matrix elements $\overline{V}_{l_{1}l_{2}l_{1}^{'}l_{2}^{'}}$
with $l_{i} = 0, x, y, z$ follow straightforwardly, but we do not write 
them in order to save space.  

The dielectric properties of the isotropic crystal in the direction 
parallel and perpendicular to the the wave vector are described by
the upper left $2 \times 2$ block, and by the third and fourth diagonal 
element in the matrix (\ref{ap6}), respectively. Note that the 
parallel and perpendicular responses with respect to any direction of the 
wave vector in the cubic crystal are identical to those along the respective
preferred
dipolar axis  $x$, and the perpendicular axes $y$ and $z$  within the model B.
Thus, all conclusions derived within the model B for the preferred direction
$x$ are valid also for any direction in the isotropic cubic crystal.
In particular the expression (\ref{d12}) for $\epsilon_M({\bf q},
\omega)$ is now valid for any orientation of ${\bf q}$.

\end{sloppypar}
\end{document}